\documentclass[reprint,
superscriptaddress,
amsmath,
amssymb,
prl,
floatfix,
english%,
]{revtex4-2}

\usepackage{tikz}
\usetikzlibrary{decorations.pathmorphing}

\usepackage{graphicx}
\usepackage{svg}
\usepackage{dcolumn}
\usepackage{bm}
\usepackage{siunitx}
\usepackage[T1]{fontenc}
\usepackage[utf8]{inputenc}
\usepackage[unicode=true, colorlinks=true, citecolor={blue!80!black}, urlcolor={blue!50!black}, linkcolor = {blue!80!black}]{hyperref}
\usepackage[normalem]{ulem}

\usepackage{physics}
\usepackage[sep=2pt, offset=1em]{simpler-wick}

\newcommand{\eps}{\varepsilon}
\newcommand{\phase}{\hat{\varphi}}
\newcommand{\Ham}{\hat{\mathcal{H}}}
\newcommand{\xqp}{x_{\rm QP}}
\newcommand{\nqp}{n_{\rm QP}}

\graphicspath{{figures/}}

\begin{document}
\widetext

\title{
Quasiparticle-induced decoherence of a driven superconducting qubit
}

\author{Mykola Kishmar}
\thanks{mk4655@columbia.edu}
\affiliation{Physics Department, Columbia University, New York, New York 10027, USA}
\author{Pavel D. Kurilovich}
\affiliation{Departments of Physics and Applied Physics, Yale University, New Haven, CT 06520, USA}
\author{Andrey Klots}
\affiliation{Google Quantum AI, Santa Barbara, CA 93117, USA}
\author{Thomas Connolly}
\affiliation{Departments of Physics and Applied Physics, Yale University, New Haven, CT 06520, USA}
\author{Igor L. Aleiner}
\affiliation{Google Quantum AI, Santa Barbara, CA 93117, USA}
\author{Vladislav D. Kurilovich}
\affiliation{Google Quantum AI, Santa Barbara, CA 93117, USA}

\begin{abstract}
We develop a theory for two quasiparticle-induced decoherence mechanisms of a driven superconducting qubit. 
In the first mechanism, an existing quasiparticle (QP) tunnels across the qubit's Josephson junction while simultaneously absorbing a qubit excitation and one (or several) photons from the drive. 
In the second mechanism, a qubit transition occurs during the non-linear absorption process converting multiple drive quanta into a pair of new QPs. 
Both mechanisms can remain significant in gap engineered qubits whose coherence is insensitive to QPs without the drive.
Our theory establishes a { {fundamental}} limitation on fidelity of the microwave qubit operations---such as readout and gates---stemming from QPs.
\end{abstract}

\maketitle

Superconducting quantum computing devices operate at cryogenic temperatures, $T \sim 10\,{\rm mK}$.
These temperatures are two orders of magnitude smaller than the superconducting energy gap $\Delta \sim 1\,{\rm K}$. 
Therefore, the number of dissociated Cooper pairs should be negligible in thermal equilibrium, $\propto \exp (-\Delta / T)$. 
In practice, though, an appreciable fraction of Cooper pairs remains broken down to the lowest accessible temperatures \cite{glazman2021, aumentado2004, shaw2008, lenander2011, riste2013, pop2014, vool2014, wang2014, bespalov2016, kyle2018, kyle2019, spencer2022, connolly2024}.
Each broken Cooper pair contributes two quasiparticle (QP) excitations  to the superconductor.
{  The density of nonequilibrium QPs observed in stationary conditions is {$\xqp = \nqp / (2\nu_0 \Delta) \sim 10^{-9} - 10^{-5}$} [the density $\nqp$ is normalized by the so-called Cooper pair density $2\nu_0 \Delta$, where $\nu_0$ is the normal-state density of states in the metal].
{Impacts of ionizing radiation with the device \cite{vepsalainen2020, wilen2021, cardani2021, martinis2021, mcewen2022resolving} can temporarily elevate $\xqp$ to values $\gtrsim 10^{-4}$~\cite{yelton_2024}.}} 

{Presence of nonequilibrium QPs opens up a channel for the decay of an excited superconducting qubit  \cite{lutchyn2005, lutchyn2006, martinis2009, catelani2011_prl, catelani2011_prb, catelani2012}. In this channel, the qubit energy is absorbed by a QP tunneling across the qubit's Josephson junction (JJ). 
{For example, due to the QP tunneling, the qubit relaxation time drops to sub-${\rm \mu s}$ scale following radiation impacts (see Fig.~4 of Ref.~\cite{mcewen2022resolving}).}

A {recently developed} strategy for counteracting the QP tunneling is known as {\it gap engineering} \cite{kamenov2024, connolly2024, mcewen2024, marchegiani2022}. It amounts to implementing qubits with a difference $\delta \Delta = \Delta_R - \Delta_L$ in gap values across the JJ, see Fig.~\ref{fig:summary}(a). 
The gap difference $\delta \Delta$ acts as a potential barrier for QPs. 
If $\delta \Delta / \hbar$ exceeds the qubit frequency $\omega_{\rm q}$---and QPs are ``cold'' \cite{spencer2022, connolly2024}---then QPs are unable to tunnel.
This blocks the QP channel of the energy relaxation.
The success of gap engineering is exemplified by a dramatic suppression of the effects of radiation impacts on qubit arrays~\cite{mcewen2024}.

Although the gap engineering reduces the QP tunneling, it neither removes QPs from the device nor prevents processes generating new QPs. 
It is natural to ask: are there ways in which QPs damage the performance of {\it gap-engineered} qubits?

To answer this question, let us note that implementation of qubit operations, such as readout or gates, relies on the irradiation of the qubit by microwaves. 
We will show in this Letter that the microwave drive can {\it reanimate} QP tunneling, and bring back the qubit decoherence.  
This happens through a process in which a qubit excitation and a microwave quantum $\hbar \omega_{\rm d}$ are simultaneously absorbed by a QP impinging on the JJ, see Fig.~\ref{fig:summary}(a). 
The QP tunneling becomes allowed above the microwave absorption threshold, $\omega_{\rm d} > \delta \Delta / \hbar - \omega_{\rm q}$. 
For a strong drive, this effect can completely negate the benefit in the coherence gained from the gap engineering. 

\begin{figure*}[t]
\centering
  \begin{center}
    \includegraphics[scale = 1]{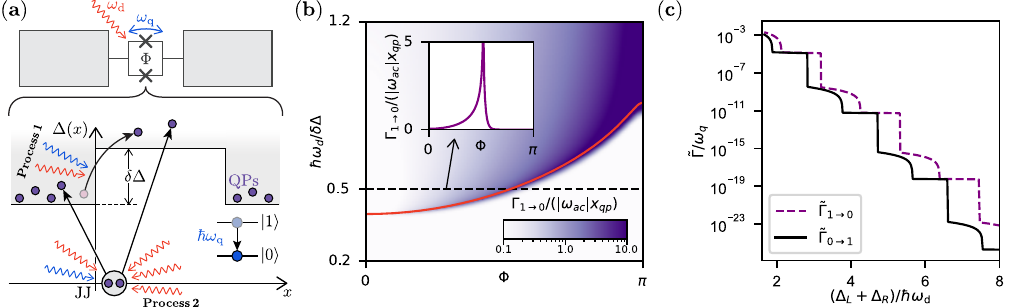}
    \caption{({\bf a}) QP-induced qubit transitions in a driven flux-tunable transmon. {\bf Process~1:} qubit relaxation is caused by the QP tunneling assisted by the absorption of one drive photon; for ``cold'' QPs, this process requires $\omega_{\rm d} + \omega_{\rm q} > \delta \Delta / \hbar$. {\bf Process~2:} qubit transition occurs when $n$ drive photons break a Cooper pair at the JJ. The process takes effect at $n \omega_{\rm d} + \omega_{\rm q} > (\Delta_L + \Delta_R) / \hbar$ [$\Delta_\alpha$ is the gap on $\alpha = L$ or $R$ side of the JJ]. ({\bf b})~Dependence of {\bf process 1} rate on flux bias and drive frequency [Eq.~\eqref{eq:1ph_10}]. Red line is the threshold, $\omega_{\rm d} + \omega_{\rm q}(\Phi) = \delta \Delta / \hbar$. Inset: $\Gamma_{1 \rightarrow 0}$ at $\omega_{\rm d} = \delta \Delta / (2 \hbar)$; the rate is normalized by $|\omega_{\rm ac}|\, x_{\rm qp}$, where $\omega_{\rm ac}$ is the ac-Stark shift. 
    ({\bf c}) Rates of the qubit transitions caused by {\bf process 2}
    [Eq.~\eqref{eq:CP_rate}]. 
    The parameters are $\Delta_L / h = 45\,{\rm GHz}$, $\Delta_R / h = 55 {\rm GHz}$, and $\omega_{\rm q}(0) / 2\pi = 6$~GHz; the transmon's SQUID is asymmetric, $E_{J1} / E_{J2} = 0.9$. In panel (b), the QP energy distribution is $n_L(\varepsilon) \propto e^{-(\varepsilon - \Delta_L) / T}$, with $T = 5 \cdot 10^{-3}$~K. In panel (c), the ac-Stark shift is $|\omega_{\rm ac}| / \omega_{\rm q} = 7\cdot 10^{-3}$.} 
    \label{fig:summary}
  \end{center}
\end{figure*}

{We develop a theory of the microwave-assisted QP tunneling for a flux-tunable transmon~\cite{Koch2007}. 
Our theory relates the rate of the QP-induced decoherence $\Gamma$ to power and frequency of microwaves, and elucidates the dependence of $\Gamma$ on flux bias $\Phi$. 
In addition to the described single-photon process, we account for possibility of QP tunneling assisted by multiple photons. 
Such processes have lower frequency thresholds.} 
On top of the processes involving the existing QPs, multi-photon processes can generate {\it new} QPs.
Indeed, $n \gtrsim (\Delta_L + \Delta_R) / (\hbar \omega_{\rm d})$ photons have sufficient energy to break a Cooper pair at the JJ.
The pair-breaking causes undesired qubit transitions as well, see Fig.~\ref{fig:summary}(a). 

Our results establish a {fundamental} limitation on the performance of microwave qubit operations imposed by the presence (and potential creation) of QPs.

{\it Model} --- To find the rates of the QP-induced decoherence, we start with the Hamiltonian of a driven flux-tunable transmon:
\begin{equation}
    \Ham = \Ham_\varphi(t) + \Ham _{\rm qp} + \Ham _{\rm T}(t) + \Ham_{\rm CP}(t). \label{eq:Hamiltonian}
\end{equation}
The first term describes the dynamics of the quantized phase difference $\phase$ across the transmon's JJs:
\begin{align}
    \Ham _\varphi(t) = 4 E_C \hat{N}^2  -  \sum_{j = 1,2} E_{Jj}\cos[\phase_j - \varphi_{\rm d}(t)], \label{eq:H_phi}
\end{align}
where $E_{Jj}$ and $4E_C$ are Josephson and charging energies, respectively, and $\hat{\varphi}_j = \hat{\varphi} - (-1)^j \pi \Phi / \Phi_0$ [index $j = 1, 2$ denotes the two junctions]. The operator $\hat{N} = - i d/ d\phase$ is canonically conjugated to $\phase$, and corresponds to the number of Cooper pairs transferred between the qubit's pads. The microwave drive is described by $\varphi_{\rm d}(t) = a\,\mathrm{cos}(\omega_{\rm d} t)$. {The drive amplitude $a$ can be related to the qubit ac-Stark shift, $\omega_{\rm ac} = -{\tfrac{1}{8}}\omega_{\rm q} (\Phi) a^2 \omega_{\rm d}^4 / [\omega_{\rm d}^2 - \omega_{\rm q}^2(\Phi)]^2$ \cite{krantz2019, blais2021}. Here, $\omega_{\rm q}(\Phi)$ is the qubit frequency in the absence of microwaves.} 

The second term in Eq.~\eqref{eq:Hamiltonian} is the Hamiltonian of QPs:
\begin{equation}
    \Ham_{\rm qp} = \sum_{k\sigma\alpha} \eps_{k\alpha} \hat{\gamma}^\dagger_{k\sigma\alpha} \hat{\gamma}_{k\sigma\alpha},
\end{equation}
where $\eps_{k\alpha} = \sqrt{\Delta_\alpha^2 + \xi^2_{k\alpha}}$ is the energy of a QP in the lead $\alpha = R$ or $L$, and $\xi_{k\alpha}$ is the energy of a normal-metal state $k$ computed with respect to the Fermi level. The fermionic operator $\hat{\gamma}_{k\sigma\alpha}$ annihilates a QP in a state $k$ in the lead~$\alpha$; $\sigma = \pm$ is a spin index.

The term $\Ham_{\rm T}(t)$ in Eq.~\eqref{eq:Hamiltonian} describes tunneling of QPs across the two junctions in the presence of a microwave drive \footnote{A similar model was recently considered in Ref.~\onlinecite{dubovitskii2024} in the context of QP effects in superconducting cat qubits.}:
\begin{align}
    \Ham_{\rm T}(t) &= \sum_{kk'\sigma j}t_{j,kk^\prime}\big(e^{\tfrac{i}{2}[\varphi_{\rm d}(t) - \phase_j]} u_{k R} u_{k' L} \nonumber \\
    &- e^{-\tfrac{i}{2}[\varphi_{\rm d}(t) - \phase_j]} v_{k R} v_{k' L}\big)\hat{\gamma}^\dagger_{k\sigma R} \Hat{\gamma}_{k'\sigma L} + \text{h.c.} \label{eq:H_T}
\end{align}
The tunneling matrix element $t_{j,kk^\prime}$ is the same matrix element that defines the Josephson energy via the Ambegaokar-Baratoff relation, $E_{Jj} = \hbar f(\Delta_L, \Delta_R) G_{Tj} / e^2$, where $G_{Tj} = 4\pi^2 \nu_0^2 |t_j|^2$ is the normal-state conductance of the junction~\footnote{We make a conventional approximation, and drop the dependence of $t_{j, kk^\prime}$ on $k$ and $k^\prime$.} and $\nu_0$ is the normal density of states in the leads. Function $f(\Delta_L, \Delta_R)$ \cite{ambegaokar-baratoff} can be approximated by $f = \pi \overline{\Delta} / 4$ in the limit $\delta \Delta \ll \overline{\Delta} \equiv (\Delta_{L} + \Delta_{R}) / 2$. The BSC coherence factors are given by
$u_{k\alpha} = \sqrt{(1 + \xi_{k\alpha}/\eps_{k\alpha})/2}$ and $v_{k\alpha} = \sqrt{(1 - \xi_{k\alpha}/\eps_{k\alpha})/2}$.

The two terms in the brackets in Eq.~\eqref{eq:H_T} correspond to two interfering channels of the QP tunneling: a particle channel ($uu$-term) and a hole channel ($vv$-term).

Finally, the last term in Eq.~\eqref{eq:Hamiltonian} describes processes of Cooper pair breaking; we will specify its form below when discussing the qubit transitions due to such processes.

{\it Microwave-assisted QP tunneling} --- 
First, we find the rate of qubit transitions due to the QP tunneling assisted by the absorption of one drive photon [process 1 in Fig.~\ref{fig:summary}].
To accomplish this task, we treat $\Ham_{\rm T}(t)$ of Eq.~\eqref{eq:H_T} as a perturbation and use it in Fermi's Golden rule. 
Initially, a QP with energy $\varepsilon_{kL}$ resides in the left lead, while the qubit is in the excited state $|1\rangle$: $\ket{i} = \hat{\gamma}^\dagger_{k\sigma L}\ket{{\rm vac}, 1}$ [$|{\rm vac}\rangle$ denotes the BCS ground state]. In the final state, $\ket{f} = \hat{\gamma}^\dagger_{k'\sigma R}\ket{{\rm vac}, 0}$, the QP has tunneled across one of the two junctions, and the qubit has relaxed its energy, $1 \rightarrow 0$. The expression for the rate of this process reads
\begin{align}\label{eq:Rate}
    \Gamma^{(1)}_{1 \rightarrow 0} =& \frac{2\pi}{\hbar}\sum_{k k^\prime \sigma j} n_L(\varepsilon_{kL})   \\
    &\times \left|{\cal M}^{1 \rightarrow 0}_{j,k k^\prime}(\Phi) \right|^2 \delta (\varepsilon_{kL} + \hbar \omega_{\rm q} (\Phi) + \hbar \omega_{\rm d} - \varepsilon_{k'R}),\nonumber 
\end{align}
where $j = 1$ or $2$ denotes the junction through which the QP has tunneled; $n_L(\varepsilon_{kL})$ is the QP distribution function on the low-gap side of the junction [we neglect the presence of QPs on the high-gap side, see Fig.~\ref{fig:summary}(a)].

\begin{figure}[t]
    \centering
    \includegraphics[width=1\linewidth]{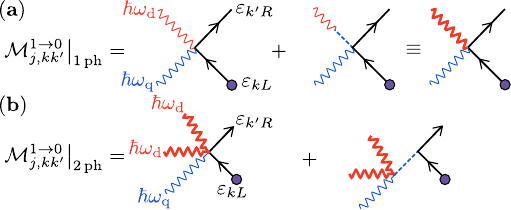}
    \caption{Diagrammatic representation of ({\bf a}) single-photon and ({\bf b}) two-photon process amplitudes. Blue and red wavy lines represent the qubit excitation and the drive, respectively; dashed blue line is the qubit response function $D_0(\omega) = \omega_{\rm q}^2 / (\omega^2 - \omega_{\rm q}^2)$. Circles and black lines represent a QP.}
    \label{fig:diagrams}
\end{figure}
To find the transition amplitude ${\cal M}$, we note that the perturbation $\varphi_{\rm d}(t)$ enters into the Hamiltonian twice, see Eqs.~\eqref{eq:H_phi} and \eqref{eq:H_T}. 
This means that there are two contributions to ${\cal M}^{1 \rightarrow 0}_{j,k k^\prime}(\Phi)$ which we depict graphically in Fig.~\ref{fig:diagrams}(a).
The first contribution describes a direct process in which the QP simultaneously absorbs the qubit excitation and the drive quantum {  [it is obtained by expanding $\exp{i[\varphi_{\rm d}(t) - \phase_j] / 2}$ in Eq.~\eqref{eq:H_T} to the first order in $\varphi_{\rm d}(t)$ and $\phase_j$]}. The second contribution accounts for the screening of the drive by the qubit's plasma oscillation. Combining the two contributions, we obtain
\begin{align} \label{eq:Matrix_Element}
    &{\cal M}^{1\rightarrow 0}_{j,kk^\prime} \big|_{\text{1 ph}} =  \frac{a}{4}\frac{\omega_{\rm d}^2\,t_j}{\omega^2_{\rm d} - \omega^2_{\rm q}(\Phi)} { \left(\frac{E_C}{\hbar \omega_{\rm q}(\Phi)}\right)^{1/2}}\\
    &\times\Bigl(u_{k L } u_{ k^\prime R}\,e^{ i(-1)^j\tfrac{\pi\Phi}{2\Phi_0} - i\frac{\varphi_{\rm m}}{2}} -v_{k L } v_{k^\prime R } e^{-i(-1)^j\tfrac{\pi\Phi}{2\Phi_0} + i\frac{\varphi_{\rm m}}{2}}\Bigr),\nonumber
\end{align}
where
$a$ is the amplitude of the field $\varphi_{\rm d}(t)$, and
\begin{equation}
    \varphi_{\rm m} = {\rm arctan}\left[\frac{E_{J2} - E_{J1}}{E_{J1} + E_{J2}} \tan \frac{\pi \Phi}{\Phi_0}\right]\notag
\end{equation}
is the minimum of the Josephson potential at flux bias~$\Phi$. 

Substituting Eq.~\eqref{eq:Matrix_Element} into Eq.~\eqref{eq:Rate}, we find:
\begin{align}
    \Gamma_{1\rightarrow 0}^{(1)} = \dfrac{|\omega_{\rm ac}|{  \xqp}}{4 \pi}&\Big\{S_+[\omega_{\rm d} + \omega_{\rm q}(\Phi)]\Big(\tfrac{\omega_{\rm q}^2(0)}{\omega_{\rm q}^2(\Phi)} - 1\Big) \label{eq:1ph_10}\\
    &+ S_{-}[\omega_{\rm d} + \omega_{\rm q}(\Phi)]\Big(\tfrac{\omega_{\rm q}^2(0)}{\omega_{\rm q}^2(\Phi)} + 1\Big) \Big\}.  \nonumber
\end{align}
Here, we expressed the power of the microwave drive in terms of the ac-Stark shift {  $\omega_{\rm ac}$ [see the discussion after Eq.~\eqref{eq:H_phi}].
The rate is proportional to the QP density $\xqp$. 
The structure factors $S_\pm[\omega]$, in turn, contain information about the energy distribution $n_L(\varepsilon)$ of QPs:
\begin{eqnarray}    
S_\pm[\omega] = \hspace{-0.85cm} \int \limits_{\max(\Delta_L,\Delta_R - \hbar    \omega)}^{+\infty}\hspace{-0.75cm}\frac{d \eps}{\overline{\Delta}}\frac{[\eps(\eps + \hbar \omega) \pm \Delta_L \Delta_R]\,n_L(\varepsilon) / \xqp}{\sqrt{\eps^2 - \Delta_L^2} \sqrt{(\eps + \hbar \omega)^2 - \Delta_R^2}}, \label{eq:QP_SF}
\end{eqnarray}
where $\overline{\Delta} = (\Delta_L + \Delta_R)/2$ and $\omega \equiv \omega_{\rm q}(\Phi) + \omega_{\rm d}$.
If $n_L(\varepsilon)$ is confined to a narrow strip $\delta E \ll \delta \Delta$ near the gap edge (i.e., the QPs are ``cold''), {then} $S_{\pm}[\omega]$ have a threshold:
\begin{equation}
\label{eq:QP_SF_asymptotics}
        S_{\pm}[\omega] = \dfrac{1}{2}\left(\frac{2 \overline{\Delta}}{\hbar\omega - \delta \Delta}\right)^{\pm 1 / 2}\Theta(\hbar\omega - \delta \Delta),
\end{equation}
where $\Theta(x)$ is a step function. 
The threshold at $\omega \equiv \omega_{\rm q}(\Phi) + \omega_{\rm d} = \delta \Delta / \hbar$ is depicted in Fig.~\ref{fig:summary}(b) by a red line.
{Eq.~\eqref{eq:QP_SF_asymptotics} applies away from the threshold's immediate vicinity, $|\hbar \omega -\delta \Delta| \gtrsim \delta E$. Close to the threshold, discontinuities are broadened by $\sim \delta E / \hbar$.
Obtaining Eq.~\eqref{eq:QP_SF_asymptotics}, we also made an assumption that $\hbar \omega - \delta \Delta \ll \overline{\Delta}$. 
In this case, $S_+[\omega] \gg S_-[\omega]$.} }

Lastly, terms in round brackets in Eq.~\eqref{eq:1ph_10} stem from the variation of the QP tunneling amplitude with the flux bias $\Phi$, cf.~Eq.~\eqref{eq:Matrix_Element}. The resulting dependence of $\Gamma^{(1)}_{1\rightarrow 0}$ on $\Phi$ deserves a comment. 
{Indeed, the largest of the two structure factors, $S_+[\omega]$, is multiplied by an expression that {\it vanishes} at $\Phi \rightarrow 0$. 
Because of this, $\Gamma^{(1)}_{1 \rightarrow 0}(\Phi)$ is {suppressed} at $\Phi = 0$, see Fig.~\ref{fig:summary}(b).
{  The suppression originates from the character of the interference between particle and hole channels of the QP tunneling.
The interference is {\it destructive} for processes in which the QP couples to the electromagnetic field an even number of times. 
In the considered process, the QP couples once to the qubit field, and once to the drive field; this leads to ${\cal M}(\Phi = 0) \propto u u - v v$ [Eq.~\eqref{eq:Matrix_Element}], where $u \approx v \approx 1 / \sqrt{2}$ for a low-energy QP.}

{\it Two-photon processes} --- Suppression of the one-photon process at $\Phi = 0$ prompts considering a relaxation process assisted by the absorption of {\it two} photons. 
{  For a two-photon process, the interference of particle and hole channels is {\it constructive}, ${\cal M}(\Phi = 0) \propto u u + v v$ [cf.~Eq.~\eqref{eq:H_T}].
Thus, despite being of a higher order in the drive power, this process} can compete with a one-photon process. For $E_J \gg E_C$, the transition amplitude is given by a sum of diagrams in Fig.~\ref{fig:diagrams}(b). Using ${\cal M}_{j,kk^\prime}^{1\rightarrow 0}|_{\rm 2\,ph}$ of Fig.~\ref{fig:diagrams}(b) in Fermi's Golden rule, we find at $\Phi = 0$:
\begin{align}    \Gamma_{1\rightarrow 0}^{(2)} &= \dfrac{1}{16 \pi} \dfrac{\omega_{\rm ac}^2 {  \xqp}}{\omega_{\rm q}(0)} \label{eq:2ph_10} \\
    &\times \left\{1 + 4 D_0[2 \omega_{\rm d} + \omega_{\rm q}(0)] \right\}^2  S_{+}[2 \omega_{\rm d} + \omega_{\rm q}(0)],\nonumber
\end{align}
where $D_0(\omega) = \omega^2_{\rm q}(0)/[\omega^2 - \omega_{\rm q}^2(0)]$ is the qubit response function.
The second term in the curly bracket originates from the parametric oscillation of the qubit frequency at $2\omega_{\rm d}$ [second diagram in Fig.~\ref{fig:diagrams}(b)].
For $\omega_{\rm d} \sim \omega_{\rm q}$, it is of the same order as the first term. According to Eq.~\eqref{eq:2ph_10}, the rate of the two-photon process is proportional to $S_+$; this needs to be contrasted with the rate of the one-photon process which is $\propto S_- \ll S_+$. 
In addition to featuring a larger structure factor, the two-photon process also remains effective in a wider range of drive frequencies, $\omega_{\rm d} > (\delta \Delta / \hbar - \omega_{\rm q})/ 2$. 
The rates of one- and two-photon processes have the same order of magnitude at $\Phi = 0$ and $\omega_{\rm ac}$ relevant for the transmon readout~\cite{supplement}.
In Ref.~\onlinecite{supplement}, we generalize Eq.~\eqref{eq:2ph_10} to arbitrary $\Phi$.

{\it Excitation process} --- Not only can QPs relax a driven qubit, they can also lead to its {\it excitation}.
To exemplify this phenomenon, we consider a $0\rightarrow 1$ qubit transition; we focus only on a one-photon process here deferring the discussion of a two-photon process to Ref.~\onlinecite{supplement}.
A direct counterpart of Eq.~\eqref{eq:1ph_10} is
\begin{align}
    \Gamma^{(1)}_{0\rightarrow 1} = \dfrac{| \omega_{\rm ac}|{  \xqp}}{4 \pi}&\Big\{S_+[\omega_{\rm d} - \omega_{\rm q}(\Phi)]\Big(\tfrac{\omega_{\rm q}^2(0)}{\omega_{\rm q}^2(\Phi)} - 1\Big) \label{eq:1ph_01}\\
    &+ S_{-}[\omega_{\rm d} - \omega_{\rm q}(\Phi)]\Big(\tfrac{\omega_{\rm q}^2(0)}{\omega_{\rm q}^2(\Phi)} + 1\Big) \Big\}, \notag
\end{align}
with structure factors given by Eqs.~(\ref{eq:QP_SF}, \ref{eq:QP_SF_asymptotics}).
Compared to the relaxation process, the excitation process has a more stringent threshold on the drive frequency; it becomes effective at $\omega_{\rm d} > \delta \Delta / \hbar + \omega_{\rm q}(\Phi)$.
We note that the excitation processes can result in ``leakage errors'' in which the qubit gets ejected into non-computational states.
For example, the rate of $1\rightarrow 2$ transition is $\Gamma^{(1)}_{1\rightarrow 2} \approx 2\Gamma^{(1)}_{0\rightarrow 1}$.

{\it Cooper-pair breaking process} ---
Qubit transitions discussed up to this point stemmed from the existing QPs in the device. 
At sufficiently high drive frequency, in addition to these processes, there appear transitions accompanied by generation of {\it new} QPs, i.e., processes in which the drive breaks Cooper pairs at the JJ. 
Provided the drive frequency satisfies $\hbar\omega_{\rm d} < \Delta_L + \Delta_R$, such processes require a simultaneous absorption of $n > 1$ photons.

The Cooper pair breaking is described by the last term in Eq.~\eqref{eq:Hamiltonian}. Explicitly, this term reads \cite{houzet2019}
\begin{align}
    &\hat{\mathcal{H}}_{\rm CP}(t) = \sum_{k k' \sigma j} \sigma\,t_{j,kk'} \big(e^{\frac{i}{2}[\varphi_{\rm d}(t) - \phase_j]} u_{k'R} v_{kL} \nonumber \\
    &+ e^{-\frac{i}{2}[\varphi_{\rm d}(t) - \phase_j]} v_{k'R}u_{kL} \big) \hat{\gamma}^\dagger_{k' \sigma R} \hat{\gamma}^\dagger_{k \overline{\sigma} L} + \text{h.c.} \nonumber \\
    &+ \sum_{j=1,2}E_{Jj}\cos[\phase_j - \varphi_{\rm d}(t)], \label{eq:HCP}
\end{align}
where $\overline{\sigma} \equiv -\sigma$ (the contribution in the third line is included to avoid double-counting the Josephson energy in Eq.~\eqref{eq:H_phi}).
The energy cost of breaking a pair and changing the state of the qubit from $\ket{i}$ to $\ket{f}$ is at least $2\overline{\Delta} + \hbar \omega_{fi}$, where  $\hbar \omega_{fi}\equiv E_{f} - E_i$ {and $\overline{\Delta} = (\Delta_L + \Delta_R) / 2$.}
The smallest number of drive quanta that can provide this energy is $n = \lceil(2\overline{\Delta} + \hbar \omega_{fi}) / (\hbar \omega_{\rm d})\rceil$ (the sign $\lceil x \rceil$ indicates the smallest integer $\geq x$).
The rate of the $n$-photon transition can be found by expanding $\Ham_{\rm CP}(t)$ in powers of $\varphi_{\rm d}(t) = a \cos (\omega_{\rm d} t)$ to order $n$, and applying Fermi's Golden rule. 
First, we consider a qubit relaxation process.
We find for its rate:
\begin{align}
    &{  \tilde{\Gamma}^{(n)}_{1\rightarrow 0} =  \frac{\omega_{\rm q}(\Phi)}{\pi\, 2^{n+1}n!^2}\left(\tfrac{|\omega_{\rm ac}|}{\omega_{\rm q}(\Phi)}\right)^n 
    \Big\{\tilde{S}_{p(n)}[\omega_{\rm q} (\Phi)+ n \omega_{\rm d}]} \label{eq:CP_rate}\\
    &{  \times \Big(\tfrac{\omega_{\rm q}^2(0)}{\omega_{\rm q}^2(\Phi)} - 1\Big)+ \tilde{S}_{-p(n)} [\omega_{\rm q} (\Phi) + n \omega_{\rm d}] \Big(\tfrac{\omega_{\rm q}^2(0)}{{\omega_{\rm q}^2(\Phi)}} +1 \Big) \Big\},} \notag
\end{align}
where $p(n)= ``+"$ if $n$ is even and $``-"$ if $n$ is odd. 
The structure factors $\tilde{S}_{\pm}[\omega]$ describing the pair breaking differ from zero only at $\omega > 2\overline{\Delta}/ \hbar$; at such frequencies, they are given by \cite{houzet2019}
\begin{align}
    \tilde{S}_\pm[\omega] = &\int \limits_{\Delta_L}^{\hbar \omega - \Delta_R} \frac{d \eps}{\overline{\Delta}} \frac{\eps(\hbar \omega - \eps) \pm \Delta_L \Delta_R}{\sqrt{\eps^2 - \Delta_L^2} \sqrt{(\hbar \omega - \eps)^2 - \Delta_R^2}}.
\end{align}
The asymptotes of $\tilde{S}_\pm[\omega]$ at $\delta \omega \equiv \omega - {  2\overline{\Delta} / \hbar \ll \overline{\Delta} / \hbar}$ are
\begin{equation}
    \label{eq:CP_SF_Asymptotics}
        \tilde{S}_+[\omega] = \pi + \dfrac{\pi}{4} \dfrac{\hbar \delta \omega}{  \overline{\Delta}}, \quad\quad \tilde{S}_-[\omega] =  \frac{\pi}{2}\dfrac
        {\hbar \delta \omega}{  \overline{\Delta}}.
\end{equation}
Equation~\eqref{eq:CP_rate} generalizes the result of Ref.~\onlinecite{houzet2019} from a single-photon case to the multi-photon one. 
{  The rate of qubit excitation, $\tilde{\Gamma}^{(n)}_{0\rightarrow 1}$, can be obtained by replacing $ \omega_{\rm q}(\Phi) \rightarrow - \omega_{\rm q}(\Phi)$ in the argument of $\tilde{S}_{\pm p(n)}$ in Eq.~\eqref{eq:CP_rate}.}

Figure~\ref{fig:summary}(c) shows the dependence of qubit excitation and relaxation rates on the drive frequency at $\Phi = 0$. 
Its main feature is that $\tilde{\Gamma}_{0 \rightarrow 1 
/ 1 \rightarrow 0}$ change exponentially with $2\overline{\Delta}/(\hbar \omega_{\rm d})$~\cite{Keldysh1964}.
The thresholds which modulate the overall exponential trend stem from switches in the number of photons required to break a Cooper pair. 
The $n$-th threshold is located at $\omega_{\rm d} = (2\overline{\Delta}/\hbar - \omega_{if})/n$.
The behavior near thresholds with even $n$ differs from that near thresholds with odd $n$. 
{This parity effect stems from the different threshold behavior of $\tilde{S}_{+}$ and $\tilde{S}_{-}$, see Eqs.~\eqref{eq:CP_rate} and \eqref{eq:CP_SF_Asymptotics}.
In Ref.~\onlinecite{supplement}, we generalize Eq.~\eqref{eq:CP_rate} to arbitrary initial and final qubit states.}

{\it Discussion and conclusions} --- In summary, we identified two new mechanisms of undesired drive-induced transitions in superconducting qubits related to QPs. In the first mechanism, an existing QP uses the energy provided by drive quanta to tunnel across the JJ and change the qubit state. The simplest process of this kind is a qubit relaxation $1 \rightarrow 0$
[see Eqs.~(\ref{eq:1ph_10},\,\ref{eq:QP_SF}) and \eqref{eq:2ph_10}]. Under more stringent conditions on the drive frequency, the qubit excitation is also possible [Eq.~\eqref{eq:1ph_01}].
In the second mechanism, the qubit transition is caused by a multi-photon absorption that breaks a Cooper pair at the junction, see Eq.~\eqref{eq:CP_rate}.
We found how the rates of these processes depend on power and frequency of the microwave tone, as well as the flux $\Phi$ controlling the qubit frequency. 
Both types of processes can occur in gap engineered qubits---that is, the qubits whose coherence is insensitive to QPs in the absence of the drive.

{  Let us establish how the studied processes impact the qubit operations. 
We start with the qubit readout.
Due to microwave-assisted QP tunneling (first mechanism), the qubit may relax during the readout pulse. 
This limits the readout fidelity.
For a pulse of duration $t_{\rm RO}$, we can bound the fidelity by \footnote{We assume $\Gamma_{1\rightarrow 0}^{(1)} \gg \Gamma_{1\rightarrow 0}^{(2)}$. This is justified for a generic $\Phi \neq 0$, provided both processes are above the respective thresholds.}
\begin{equation}\label{eq:fidelity}
1 - F \gtrsim \Gamma^{(1)}_{1 \rightarrow 0} t_{\rm RO} = \alpha\cdot |\omega_{\rm ac}| t_{\rm RO} \cdot \xqp 
\end{equation}
where $\alpha$ is a dimensionless number that depends on $\omega_{\rm q}(\Phi)$, $\omega_{\rm d}$, and the superconducting gap values $\Delta_L$ and $\Delta_R$ [see Eq.~\eqref{eq:1ph_10}]. 
The parameter $|\omega_{\rm ac}| t_{\rm RO}$ in Eq.~\eqref{eq:fidelity} is, in fact, the same parameter that determines the readout signal-to-noise ratio (SNR) \cite{clerk2010}. The requirement of high SNR fixes $|\omega_{\rm ac}| t_{\rm RO} \sim 100$ for state-of-the-art readout \cite{swiadek2024, kurilovich2025highfrequency}. 
Taking $\alpha \sim 1$---which is appropriate at a generic value of $\Phi \neq 0$---we obtain $1 - F \gtrsim 100\cdot \xqp$.
The QP density $x_{\rm QP}$ depends on experimental conditions, and varies in a wide interval. 
After the impacts of ionizing radiation \cite{mcewen2024}, it can reach values exceeding $\sim 10^{-4}$~\cite{yelton_2024}. 
This corresponds to $1 - F \gtrsim 0.01$.}

{  Similarly, the bound on the fidelity of a single-qubit gate can be represented as $1 - F \gtrsim \Gamma^{(1)}_{1 \rightarrow 0}t_{\rm gate}$, where $t_{\rm gate}$ is the gate duration. We note that the derived expression for $\Gamma^{(1)}_{1 \rightarrow 0}$ diverges for the resonant pulse [cf.~Eq.~\eqref{eq:Matrix_Element}]. The divergence should be regularized by the qubit nonlinearlity, $\omega_{\rm d} - \omega_{\rm q} \rightarrow E_C / \hbar$.
A detailed calculation yields $1 - F \gtrsim \beta \cdot \xqp / (E_C t_{\rm gate} / \hbar)$ with $\beta \sim 1$. {Parameter $E_C t_{\rm gate} / \hbar$ is close to $1$ for a gate with an optimized pulse shape \cite{chow2010}.}

The microwave-assisted QP tunneling can be suppressed at the hardware level.
For example, the use of junctions with larger $\delta \Delta$ can set processes with small photon number below their thresholds.
Another option is to implement QP traps that would keep the QPs away from the qubit's JJs (see, e.g., Ref.~\onlinecite{catelani2019}).

Finally, let us comment on the pair-breaking process (second mechanism). This process represents a limitation on the performance of a \textit{high-frequency} dispersive readout \cite{kurilovich2025highfrequency}. 
Ref.~\onlinecite{kurilovich2025highfrequency} showed that the readout fidelity can be improved by using a measurement tone with $\omega_{\rm d} \gg \omega_{\rm q}$. 
Our theory shows that $\omega_{\rm d}$ cannot be increased indefinitely; once the frequency becomes too high, the pair-breaking process starts to degrade the readout fidelity.
Consider, for example, a $5~\rm GHz$ aluminum qubit that is read out at $\omega_{\rm d} / 2\pi = 60~\rm GHz$ [this is a scaled-up version of Ref.~\onlinecite{kurilovich2025highfrequency} setup].
The gap of aluminum is $\approx 50\,{\rm GHz}$; therefore, absorption of two photons can break a Cooper pair. 
Using Eq.~\eqref{eq:CP_rate} for $n = 2$, we estimate that the fidelity is limited to $1 - F \gtrsim 0.05$. 

The mechanisms we uncover are relevant for applications of microwave drives to superconducting circuits beyond transmon qubits \cite{fluxonium2009, leghtas2015, grimm2020, frozonium2025}.  

{\it Note} --- Similar QP-induced decoherence mechanisms are also considered numerically in Ref.~\onlinecite{shoumik}, which we became aware of during the preparation of the manuscript.
Our results agree with each other. 

{\it Acknowledgments} --- We acknowledge very helpful discussions with Leonid Glazman, Alex Opremcak, Shoumik Chowdhury, Max Hays, Valla Fatemi, and Michel Devoret. 

The research of P.D.K.~and T.C.~was sponsored by the Army Research Office (ARO) under grants no.W911NF-22-1-0053 and W911NF-23-1-0051, by DARPA under grant no. HR0011-24-2-0346, by the U.S. Department of Energy (DoE), Office of Science, National Quantum Information Science Research Centers, Co-design Center for Quantum Advantage (C2QA) under contract number DE-SC0012704. The views and conclusions contained in this document are those of the authors and should not be interpreted as representing the official policies, either expressed or implied, of the ARO, DARPA, DoE, or the US Government. The US Government is authorized to reproduce and distribute reprints for Government purposes notwithstanding any copyright notation herein.

\bibliography{references}
\bibliographystyle{apsrev4-1}

\end{document}

\widetext

\title{
Supplemental Materials for\\ ``Quasiparticle-induced decoherence of a driven superconducting qubit''
}
\author{Mykola Kishmar}
\affiliation{Physics Department, Columbia University, New York, New York 10027, USA}
\author{Pavel D. Kurilovich}
\affiliation{Departments of Physics and Applied Physics, Yale University, New Haven, CT 06520, USA}
\author{Andrey Klots}
\affiliation{Google Quantum AI, Santa Barbara, CA 93117, USA}
\author{Thomas Connolly}
\affiliation{Departments of Physics and Applied Physics, Yale University, New Haven, CT 06520, USA}
\author{Igor L. Aleiner}
\affiliation{Google Quantum AI, Santa Barbara, CA 93117, USA}
\author{Vladislav D. Kurilovich}
\affiliation{Google Quantum AI, Santa Barbara, CA 93117, USA}

\maketitle
\onecolumngrid

\section{GENERAL PERTURBATION THEORY FOR THE QP TUNNELING AMPLITUDES}

In the main text, we presented derivations of rates of simplest non-linear processes, such as QP-induced qubit relaxation assisted by one or two drive photons.
The goal of this Supplement is to develop a general formalism that would allow one to compute the rates of processes involving an arbitrary number of photons. 
The problem has two non-trivial aspects. First, the drive $\varphi_{\rm d}(t)$ enters the Hamiltonian of the system twice: in the qubit term, $\Ham_\varphi(t)$, and in the QP tunneling term, $\Ham_{\rm T}(t)$. Second, both of these terms depend on  $\varphi_{\rm d}(t)$ in a non-linear way.
Because of this, each transition amplitude contains a number of contributions that interfere with one another [see, e.g., Fig.~2 of the main text]; the more photons are absorbed in the transition process, the larger the number of contributions is. 
Here, we present a perturbative, diagrammatic approach which streamlines evaluation of different contributions.

The first step in developing the perturbation theory is to specify the unperturbed exactly solvable Hamiltonian $\Ham_0$, and to define the perturbation $\Ham_{\rm int}$.
In our work, we focus on a transmon qubit, $E_J \gg E_C$.
Because the low-energy part of the transmon spectrum is close to the spectrum of a harmonic oscillator \cite{Koch2007}, it is convenient to include in $\Ham_0$ the harmonic part of the qubit Hamiltonian $\Ham_\varphi$ [see Eq.~(2) of the main text]. We also include the quasiparticle Hamiltonian $\Ham_{\rm qp}$ [Eq.~(3) of the main text]. In total,
\begin{equation}
        \Ham_0 =  4E_C \hat{N}^2 + \frac{J_0(a) E_J(\Phi) \phase^2}{2} + \sum_{k\sigma\alpha} \eps_{k\alpha} \hat{\gamma}^\dagger_{k\sigma\alpha} \hat{\gamma}_{k\sigma\alpha},
        \label{eq:H0} 
\end{equation}
where $E_J(\Phi) = \sqrt{E_{J1}^2 + E_{J2}^2 + 2 E_{J1} E_{J2} \cos (\pi \Phi / \Phi_0)}$, with $\Phi_0$ the superconducting flux quantum. The Bessel function factor $J_0(a)$ describes renormalization of the Josephson potential by the drive $\varphi_{\rm d} (t) = a \cos(\omega_{\rm d} t)$; it stems from the time-averaging relation $\overline{\cos(a\cos (\omega_{\rm d} t))} = J_0(a)$. 
We include all terms of the full Hamiltonian $\Ham$ except for the ones in Eq.~\eqref{eq:H0} in $\Ham_{\rm int}$.

Due to the system non-linearity, in addition to the oscillation at $\omega_{\rm d}$, the interaction Hamiltonian $\Ham_{\rm int}$ oscillates at the overtones of the drive frequency, $e^{-in\omega_{\rm d} t}$. 
To account for this feature, we perform the Fourier decomposition of $\Ham_{\rm int}$.
Simultaneously, we also expand $\Ham_{\rm int}$ into powers $\phase^m$ of the qubit phase. 
Such a decomposition systemizes all elementary interaction processes involving the qubit, the drive, and the quasiparticles.
The decomposition reads:
\begin{equation}\label{eq:H_int}
    \Ham_{\rm int} = \sum_{m > 2} \Ham_\varphi^{(m, 0)} + \sum_{n \neq 0} \sum_{m \geq  1} e^{-in\omega_{\rm d} t}\Ham_\varphi^{(m, n)} + \sum_{j, n, m} e^{-in\omega_{\rm d} t}\Ham_{{\rm T}j}^{(m, n)}.
\end{equation}
where 
\begin{align}
    \Ham_\varphi^{(m,n)}\label{eq:Hphase_coeffs} =  -i^n(-i)^m \frac{a^n}{2^n n!}\dfrac{\phase^m}{m!} \dfrac{1 + (-1)^{m+n}}{2} Z_n(a)E_J(\Phi),
\end{align}
and
\begin{align}
        \Ham_{{\rm T}j}^{(m,n)} =& i^n(-i)^m \frac{a^n}{4^n n!}\dfrac{\phase^m}{2^m m!}\notag \\
        &\times \sum_{k k'\sigma} Z_n(a / 2) t_j  \Big( u_{k R} u_{k' L} e^{i(-1)^j \frac{\pi \Phi}{2 \Phi_0} - i\frac{\varphi_m}{2}} - (-1)^{m+n}  v_{k R}v_{k' L} e^{-i(-1)^j \frac{\pi \Phi}{2 \Phi_0} + i\frac{\varphi_m}{2}} \Big) \hat{\gamma}^\dagger_{k \sigma R} \hat{\gamma}_{k' \sigma L} + \text{h.c.}  \label{eq:HT_coeffs}
    \end{align}
Here, index $j = 1,2$ labels the two Josephson junctions.
{In Eqs.~\eqref{eq:Hphase_coeffs} and \eqref{eq:HT_coeffs}, we introduced factors $Z_n(a) = J_n(a) 2^n n! / a^n$, where $J_n(a)$ is the Bessel function and $a$ is the amplitude of the drive [in the derivation, we used $\exp(i q \cos z) = \sum_n i^n e^{i n z} J_n(q)$]. These factors satisfy $Z_n(0) = 1$; they describe renormalization the $n$-photon process amplitudes in the case of the strong drive, $a \gtrsim 1$. }
In the present discussion, we focus on tunneling of the existing QPs; this is why we dispensed with the Cooper-pair breaking terms in Eq.~\eqref{eq:H_int}.

\begin{figure*}[t]
    \includegraphics[width = 1\textwidth]{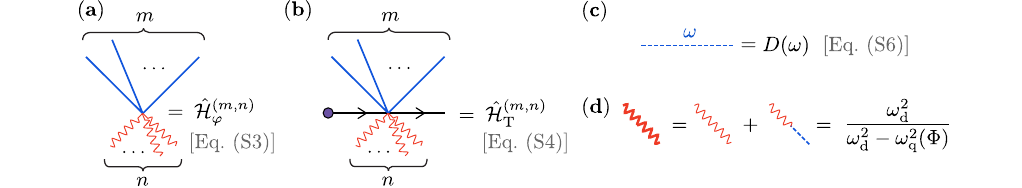}
    \caption{ \label{fig:vertices}
    Graphical representation of the diagrammatic elements: ({\bf a}) the vertex describing the qubit non-linearity and interaction of the qubit with the drive, ({\bf b}) the QP tunneling vertex, ({\bf c}) the qubit propagator, and ({\bf d}) definition of a bold drive line.}
\end{figure*}

With $\Ham_0$ and $\Ham_{\rm int}$ defined, we can find transition amplitudes in a standard way by evaluating matrix elements of the evolution operator in the interaction representation \cite{ziman}. 
Same as in the main text, here we will focus on the processes of the first order in the QP tunneling matrix element $t_j$. 
The amplitude of a transition in which a QP tunnels from a state $k$ in the left lead to the state $k^\prime$ in the right lead and changes the qubit state $i \rightarrow f$ can be represented as
\begin{align}
     \bra{\text{vac},f}\hat{\gamma}_{k' \sigma R} \mathcal{T} e^{- i \int_{-\infty}^{+\infty} dt \Ham_{\rm int}(t)} \hat{\gamma}^\dagger_{k \sigma L}\ket{\text{vac},i}_{\rm c} = - i \sum_n \mathcal{M}_{j,kk'}^{i \rightarrow f}(\Phi) \Big|_{n\,{\rm ph}} \times 2 \pi \delta(\eps_{kL} + n \cdot \hbar \omega_{\rm d} - \hbar \omega_{fi} - \eps_{k'R}), \label{eq:texp}
\end{align} 
where $\Ham_{\rm int}(t) = e^{i\Ham_0 t} \Ham_{\rm int} e^{-i\Ham_0 t}$, $\mathcal{T}$ denotes the time ordering, and ``vac'' labels the BCS vacuum.
Individual amplitudes $\mathcal{M}_{j,kk'}^{i \rightarrow f}(\Phi)|_{n\,{\rm ph}}$ can be found by expanding the exponent, and grouping together terms that oscillate with $t$ at the frequency of the $n$-photon transition [see the argument of the delta-function]; only the connected processes need to be considered \cite{ziman}, as indicated by the subscript ``c''. To automate this procedure, we use Eqs.~\eqref{eq:H0}--\eqref{eq:HT_coeffs} to introduce diagrammatic rules for the amplitudes.

The rules are summarized in Fig.~\ref{fig:vertices}. 
There are two types of interaction vertices.
The vertices of type $\Ham_\varphi^{(m, n)}$ correspond to the interaction between the qubit and the drive; the indices $m$ and $n$ determine, respectively, a number of qubit lines and a number of drive photon lines entering the vertex. 
The factor $1+(-1)^{m+n}$ in Eq.~\eqref{eq:Hphase_coeffs} indicates that the only nonzero vertices of type $\Ham_\varphi^{(m,n)}$ are those containing an even number of lines (i.e., $m+n$ is even).
The vertices of type $\Ham_{\rm T}^{(m, n)}$ describe the coupling of the tunneling QP to the qubit and the drive.
In diagrams, vertices are connected by retarded qubit propagators 
\begin{align}
    D(\omega) = \varphi_{\rm ZPF}^2 \frac{2 \omega_{\rm q}(\Phi)}{\omega^2 - \omega_{\rm q}^2(\Phi)},
    \label{eq:Propagator}
\end{align}
where {$\varphi_{\rm ZPF} = (2E_C / E_J)^{1/4}$ is the zero-point fluctuation of the qubit phase} [in the main text, we also use the dimensionless ``qubit response function'' $D_0(\omega) = E_J D(\omega)$, see Eq.~(11)]. 
We depict the qubit propagator by a dashed blue line [see Fig.~\ref{fig:vertices}(c)].
Lastly, we illustrate initial and final states of the qubit by wavy blue lines; each such line brings about a factor of $\varphi_{\rm ZPF}$ to the amplitude. Let us now present two examples of the application of the diagram rules.

\subsection*{Example 1: Qubit relaxation assisted by one drive photon}
To illustrate application of the general diagrammatic rules, we first provide a detailed derivation of Eq.~(6) of the main text for ${\cal M}_{j,kk^\prime}^{1\rightarrow 0}|_{1{\rm ph}}$.
There are two diagrams contributing to this amplitude [see Fig.~1(a) of the main text].
The first diagram contains the vertex $\Ham_{\rm T}^{(1,1)}$.
The second diagram contains vertices $\Ham_{\rm T}^{(2,0)}$ and $\Ham_\varphi^{(1,1)}$ connected by the qubit propagator $D(\omega_{\rm d})$. 
Using Eqs.~\eqref{eq:Hphase_coeffs} and \eqref{eq:HT_coeffs}, we obtain
\begin{align}
    \mathcal{M}^{1 \rightarrow 0}_{j,kk'} \Big|_{\text{1 ph}} &= \varphi_{\rm ZPF}t_j \frac{a}{8} \left(u_{kR}u_{k'L}e^{i(-1)^j\frac{\pi \Phi}{2 \Phi_0} - i \frac{\varphi_m}{2}} - v_{kR}v_{k'L}e^{-i(-1)^j\frac{\pi \Phi}{2 \Phi_0} + i \frac{\varphi_m}{2}} \right) \nonumber\\
    & +\varphi_{\rm ZPF} t_j\frac{a}{8} E_J D(\omega_{\rm d}) \left(u_{kR}u_{k'L}e^{i(-1)^j\frac{\pi \Phi}{2 \Phi_0} - i \frac{\varphi_m}{2}} - v_{kR}v_{k'L}e^{-i(-1)^j\frac{\pi \Phi}{2 \Phi_0} + i \frac{\varphi_m}{2}} \right)\notag \\
    &= \varphi_{\rm ZPF} t_j\frac{a}{8} [1 + E_J D(\omega_{\rm d})] \left(u_{kR}u_{k'L}e^{i(-1)^j\frac{\pi \Phi}{2 \Phi_0} - i \frac{\varphi_m}{2}} - v_{kR}v_{k'L}e^{-i(-1)^j\frac{\pi \Phi}{2 \Phi_0} + i \frac{\varphi_m}{2}} \right),\label{eq:M10_1ph}
\end{align}
where we assumed $a \ll 1$ and expanded the Bessel function $J_1(a)$ to the first order in $a$.  
The second term in the square bracket accounts for the screening of the drive by the qubit. Effectively, it leads to renormalization of the drive amplitude to 
\begin{align}\label{eq:renorm}
    \tilde{a} = a\left[1 + E_J D(\omega_{\rm d}) \right] = a \frac{\omega_{\rm d}^2}{ \omega_{\rm d}^2 - \omega_{\rm q}^2(\Phi)}
\end{align}
[we used Eq.~\eqref{eq:Propagator} with $\varphi_{\rm ZPF} = (\omega_{\rm q} / 2E_J)^{1/2}$]. Combination of Eqs.~\eqref{eq:M10_1ph} and \eqref{eq:renorm} constitutes Eq.~(6) of the main text.

In passing, we note that renormalization of the drive amplitude occurs not only in single-photon process, but in multi-photon processes too. We represent the renormalization graphically by bold wavy lines entering the vertices, see Fig.~\ref{fig:vertices}(d). Each bold line contributes a factor $\omega_{\rm d}^2 / [\omega_{\rm d}^2 - \omega_{\rm q}^2(\Phi)]$ to the diagram.

\subsection*{Example 2: qubit relaxation due to the absorption of three and four drive photons} 

\begin{figure*}[h]
    \includegraphics[width = 1\textwidth]{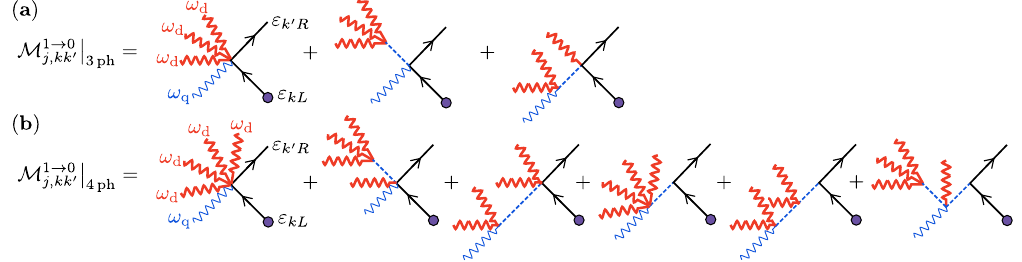}
    \caption{ \label{fig:3_ph_4_ph_diagrams} Diagrams for amplitudes of the qubit relaxation processes in which the tunneling quasiparticle absorbs ({\bf a}) three photons and ({\bf b}) four photons.}
\end{figure*}

The diagrammatic rules formulated in Fig.~\ref{fig:vertices} and Eqs.~\eqref{eq:Hphase_coeffs}, \eqref{eq:HT_coeffs}, \eqref{eq:Propagator}, and \eqref{eq:renorm} allow one to compute amplitudes of processes involving any number of photons. 
To provide an example that goes beyond the main text, in Fig.~\ref{fig:3_ph_4_ph_diagrams} we present diagrams for $\mathcal{M}^{1 \rightarrow 0}_{j,kk'} |_{\text{3 ph}}$ and $\mathcal{M}^{1 \rightarrow 0}_{j,kk'} |_{\text{4 ph}}$:\\

\noindent We refrain from specifying expressions for the amplitudes explicitly; these expressions are bulky, and follow straightforwardly from the rules in Fig.~\ref{fig:vertices}. Note that we have not included diagrams with closed qubit loops in Fig.~\ref{fig:3_ph_4_ph_diagrams}. Such diagramms have a meaning of additional screening due to the fluctuations in non-linear system, and they are suppressed by a parameter $E_{C} / E_J \ll 1$.

\section{Generalization of Eq.~(10) to arbitrary flux $\Phi$}
In the main text, we computed the rate of  qubit relaxation due to the two-photon-assisted QP tunneling at $\Phi = 0$. Here, we generalize the result to arbitrary flux $\Phi$.

\begin{figure*}[t]
\centering
  \begin{center}
    \includegraphics[scale = 1]{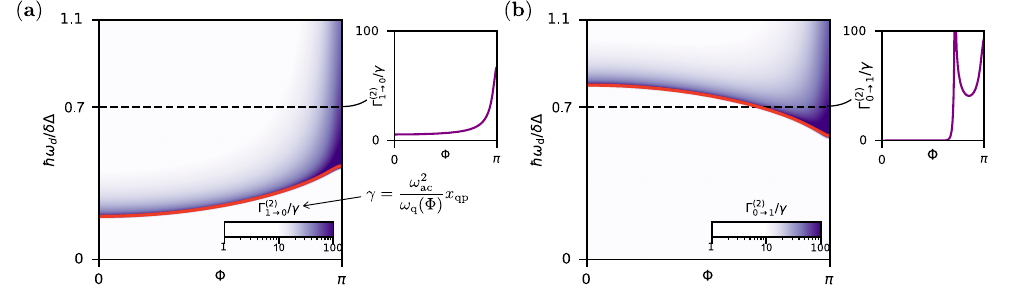}
    \caption{({\bf a}) \label{fig:Fig1_sup} Dependence of $\Gamma^{(2)}_{1 \rightarrow 0}$ on flux bias and drive frequency [Eq.~\eqref{eq:Gamma_10_2ph}]. Under the assumption of cold QPs, the process is active only above the threshold, $\omega_{\rm d} > \omega_{\rm th} = (\delta \Delta/\hbar - \omega_{\rm q}) / 2$ [red line]. Inset: $\Gamma^{(2)}_{1 \rightarrow 0}$ at $\omega_{\rm d} = 0.7 \cdot \delta \Delta/\hbar$; the rate is normalized by $\gamma = \omega_{\rm ac}^2 x_{\rm qp}/\omega_q(\Phi)$. ({\bf b}) Rate of the two-photon qubit excitation $\Gamma^{(2)}_{0 \rightarrow 1}$ [Eq.~\eqref{eq:Gamma_12_2ph}]. The threshold for this process is $\omega_{\rm th} = (\delta \Delta/\hbar + \omega_{\rm q} ) / 2$. The inset shows $\Gamma^{(2)}_{1 \rightarrow 0}$ at $\omega_{\rm d} = 0.7 \cdot \delta \Delta/\hbar$. The additional peak stems from the divergence at $2\omega_{\rm q}(\Phi) \approx 2 \omega_{\rm d}$ (see discussion around Eq.~\eqref{eq:divergence} for details). The parameters used here are $\delta \Delta = 2 \pi \cdot 10$~GHz, $\overline{\Delta} = 2 \pi \cdot 50$~GHz, $\omega_{\rm q}(0) = 2 \pi \cdot 6$~GHz, and the temperature $T = 5 \cdot 10^{-3}$~K. The transmon's SQUID is assumed to be asymmetric, with $E_{J1} / E_{J2} = 0.9$.}
    \end{center}
\end{figure*}

The amplitude of the two-photon process is given by the two diagrams depicted in Fig.~2(b) of the main text. At arbitrary $\Phi$, the expression for the amplitude reads:
\begin{align}
   \mathcal{M}^{1 \rightarrow 0}_{j,kk'} \Big|_{\text{2 ph}} = &- i\varphi_{\rm ZPF} \frac{a^2}{64}\left[1 + E_J D(\omega_{\rm d}) \right]^2  t_j\left(u_{kR}u_{k'L}e^{i(-1)^j\frac{\pi \Phi}{2 \Phi_0} - i \frac{\varphi_m}{2}} + v_{kR}v_{k'L}e^{-i(-1)^j\frac{\pi \Phi}{2 \Phi_0} + i \frac{\varphi_m}{2}} \right) \label{eq:M_2ph} \\
    - & i \varphi_{\rm ZPF} \frac{a^2}{16} \left[1 + E_J D(\omega_{\rm d}) \right]^2 t_j E_J D(2 \omega_{\rm d} + \omega_{\rm q}(\Phi))  \left(u_{kR}u_{k'L}e^{i(-1)^j\frac{\pi \Phi}{2 \Phi_0} - i \frac{\varphi_m}{2}} + v_{kR}v_{k'L}e^{-i(-1)^j\frac{\pi \Phi}{2 \Phi_0} + i \frac{\varphi_m}{2}} \right). \notag
\end{align}
Using $\mathcal{M}^{1 \rightarrow 0}_{j,kk'} |_{\text{2 ph}}$ in Fermi's Golden rule, we obtain for the rate [compare to Eq.~(10) of the main text]:
\begin{align}
    \Gamma^{(2)}_{1 \rightarrow 0} = \frac{1}{32\pi} \frac{\omega_{\rm ac}^2}{\omega_{\rm q}(\Phi)}& \left\{1 + 4 E_J D[2 \omega_{\rm d} + \omega_{\rm q}(\Phi)] \right\}^2 \notag \\
    &\times \Bigg\{S_+[2 \omega_{\rm d} + \omega_{\rm q}(\Phi)]\left( \frac{\omega_{\rm q}^2(0)}{\omega_{\rm q}^2(\Phi)} + 1 \right) + S_-[2 \omega_{\rm d} +  \omega_{\rm q}(\Phi)]\left( \frac{\omega_{\rm q}^2(0)}{\omega_{\rm q}^2(\Phi)} - 1 \right)\Bigg\}.\label{eq:Gamma_10_2ph}
\end{align}
The structure factors $S_\pm$ are given by Eq.~(8) of the main text. 
If the QPs are ``cold'' [see the discussion around~Eq.~(9)], then the two-photon process becomes active only above the threshold, $\omega_{\rm d} > [\delta \Delta / \hbar - \omega_{\rm q}(\Phi)] / 2$. 
The dependence of Eq.~\eqref{eq:Gamma_10_2ph} on $\Phi$ and $\omega_{\rm d}$ is illustrated in Fig.~\ref{fig:Fig1_sup}(a).

\section{Comparison of one- and two-photon qubit relaxation processes at $\Phi = 0$}
The interference between particle and hole channels of the QP tunneling results in a non-trivial comparison between one- and two-photon-assisted qubit relaxation processes at $\Phi = 0$. 
For the one-photon process, the interference is destructive, and the rate $\Gamma^{(1)}_{1 \rightarrow 0}\propto x_{\rm QP} |\omega_{\rm ac}| S_-$.
For the two-photon process, the interference is constructive, and the rate is determined by a much larger structure factor, $S_+ \gg S_-$. 
In the main text, we showed $\Gamma^{(2)}_{1 \rightarrow 0}\propto x_{\rm QP} [\omega^2_{\rm ac} / \omega_{\rm q}(0)] S_+$.
Therefore, even though the rate of the two-photon process is suppressed by an extra factor of $|\omega_{\rm ac}|/\omega_{\rm q}$, it can compete with the rate of the one-photon process.
Here, we compare the two rates for the parameters corresponding to state-of-the-art readout.

We use equations (7), (9), and (10) of the main text to evaluate the ratio $\Gamma^{(2)}_{1 \rightarrow 0}/\Gamma^{(1)}_{1 \rightarrow 0}$. We assume that $\hbar \omega_{\rm d} > \delta \Delta - \hbar  \omega_{\rm q}(0)$. In this case, both processes are effective. The results for the ratio of the rates is
\begin{align}
    \frac{\Gamma^{(2)}_{1 \rightarrow 0}}{\Gamma^{(1)}_{1 \rightarrow 0}} = \frac{1}{8} \frac{|\omega_{\rm ac}|}{ \omega_{\rm q}(0)} \frac{2 \overline{\Delta}}{\sqrt{\hbar [\omega_{\rm d} + \omega_{\rm q}(0)] - \delta \Delta}\sqrt{\hbar [2\omega_{\rm d} + \omega_{\rm q}(0)] - \delta \Delta}} \Bigg\{1 +  \frac{4 \omega_{\rm q}^2(0)}{[2 \omega_{\rm d} + \omega_{\rm q}(0)]^2 - \omega_{\rm q}^2(0)}\Bigg\}^2.
\end{align}
For the estimate, let us first use the parameters of Ref.~\onlinecite{swiadek2024}, i.e., $\omega_{\rm q}/2\pi = 5.6\,{\rm GHz}$, the readout frequency $\omega_{\rm d} / 2\pi = 6.9\,{\rm GHz}$, and $|\omega_{\rm ac}| / \omega_{\rm q} \approx 0.02$.
In addition to these parameters, the estimate depends on the closeness of the drive frequency to the absorption threshold; we will take $\omega_{\rm d} + \omega_{\rm q} - \delta\Delta / \hbar = 2\pi \cdot 1{\rm GHz}$. In this case, we obtain $\Gamma_{1\rightarrow 0}^{(2)} / \Gamma_{1\rightarrow 0}^{(1)} \approx 0.15$. 
A similar estimate for the parameters of  Ref.~\onlinecite{kurilovich2025highfrequency} yields $\Gamma_{1\rightarrow 0}^{(2)} / \Gamma_{1\rightarrow 0}^{(1)} \approx 0.4$. Thus, the rates of one- and two-photon processes are comparable at $\Phi = 0$.

\section{Two-photon process of qubit excitation} \label{sec:Nonlinear}
As we showed in the main text, in addition to qubit relaxation, QP tunneling can result in qubit {excitation}. A single-photon excitation process has a stringent frequency threshold taking place at $\omega_{\rm d} > \delta \Delta/\hbar + \omega_{\rm q}(\Phi)$ [see Eq.~(12) of the main text].
This motivates a consideration of multi-photon excitation processes, which have lower thresholds.

As an example, here we consider a process of $0\rightarrow 1$ qubit transition accompanied by absorption of two photons by a QP. The calculation is identical to the derivation of Eq.~(\ref{eq:Gamma_10_2ph}) and results in
    \begin{align}
        \Gamma^{(2)}_{0 \rightarrow 1}  = \frac{1}{32\pi} \frac{\omega_{\rm ac}^2}{\omega_{\rm q}(\Phi)}& \left\{1 + 4 E_J D[2 \omega_{\rm d} - \omega_{\rm q}(\Phi)] \right\}^2  \nonumber \\
        &\times \Bigg\{S_+[2 \omega_{\rm d} - \omega_{\rm q}(\Phi)]\left( \frac{\omega_{\rm q}^2(0)}{\omega_{\rm q}^2(\Phi)} + 1 \right) + S_-[2 \omega_{\rm d} -  \omega_{\rm q}(\Phi)]\left( \frac{\omega_{\rm q}^2(0)}{\omega_{\rm q}^2(\Phi)} - 1 \right)\Bigg\}. \label{eq:Gamma_12_2ph}
    \end{align}
    Such process becomes active when $\omega_{\rm d} > (\delta \Delta/\hbar + \omega_{\rm q}(\Phi)) / 2$. An interesting feature of Eq.~\eqref{eq:Gamma_12_2ph} is the divergence of the transition rate at $2\omega_{\rm d} \approx 2 \omega_{\rm q}(\Phi)$. To make it explicit, we rewrite the propagator $D(2\omega_{\rm d} - \omega_{\rm q}(\Phi))$ as
    \begin{align}\label{eq:divergence}
        E_J D[2 \omega_{\rm d} - \omega_{\rm q}(\Phi)] = \frac{\omega_{\rm q}^2(\Phi)}{2 \omega_{\rm d} [2\omega_{\rm d} - 2\omega_{\rm q}(\Phi)] }.
    \end{align} 
    This divergence stems from a resonance involving a virtual $0\rightarrow 2$ transition, $2\omega_{\rm d} = \omega_{02}$ \footnote{
The accuracy with which Eq.~\eqref{eq:Gamma_12_2ph} was derived does not capture the difference between $\omega_{02}$ and $2 \omega_{\rm q}$; this difference is $\sim E_C$ and is beyond the scope of tree-level diagrams.
}. Close to the resonance, the rate of $0 \rightarrow 1$ transition  exceeds the rate of the two-photon-assisted qubit relaxation,
\begin{align}
\frac{\Gamma^{(2)}_{0\rightarrow 1}}{\Gamma^{(2)}_{1 \rightarrow 0}} \sim \left(\frac{\omega_{\rm d}}{\omega_{\rm d} - \omega_{02}/2}\right)^2 \gg 1
\end{align}
(provided both processes are allowed by the energy conservation). The dependence of Eq.~\eqref{eq:Gamma_12_2ph} on the flux bias and the drive frequency is demonstrated on Fig.~\ref{fig:Fig1_sup}(b).

\section{Generalization of Eq.~(13) to arbitrary initial and final qubit states}
In the main text, we have found the rate of $1 \rightarrow 0$ qubit transition due to the Cooper-pair breaking processes [see Eq.~(13) of the main text]. 
Here, we generalize that expression to arbitrary initial ($|i\rangle$) and final ($|f\rangle$) qubit states. 

To separate processes involving absorption of different number of photons, we begin by decomposing the $\Ham_{\rm CP}(t)$ defined in Eq.~(12) into Fourier series:
\begin{align}
    \Ham_{\rm CP}(t) = \sum_{n \neq 0} e^{-in \omega_{\rm d} t} \Ham_{\rm CP}^{(n)} + \text{h.c.}
\end{align}
Here, the coefficients $\Ham_{\rm CP}^{(n)}$ are given by
\begin{align}
    \Ham_{\rm CP}^{(n)} = i^n \frac{a^n}{4^n n!} \sum_{k k' \sigma j}\sigma Z_n(a / 2) t_{j,kk'}\Big( e^{- \frac{i}{2} \phase_j} u_{k' R} v_{k L} + (-1)^n e^{\frac{i}{2} \phase_j} v_{k' R}u_{k L}\Big) \hat{\gamma}^\dagger_{k' \sigma R} \hat{\gamma}^\dagger_{k \overline{\sigma} L},
\end{align}
where $Z_n(a / 2)$ is a renormalization factor satisfying $Z_n(0) = 1$ [see the discussion after Eq.~\eqref{eq:HT_coeffs} for the definition], and $a$ is the amplitude of the drive.
Using the Fourier decomposition in Fermi's Golden rule, we find the transition rate stemming from the $n$-photon Cooper pair breaking process:
\begin{align}\label{eq:cp_breaking_any_states}
    \tilde{\Gamma}^{(n)}_{i \rightarrow f} = \frac{8}{\pi} \sum_{j = 1,2} \frac{E_{Jj}}{\hbar} \left(\frac{a^n}{4^n n!} \right)^2 \Big\{ \tilde{S}_{p(n)}[\omega_{if} + n \omega_{\rm d}] |\bra{f} \cos(\phase_j/2) \ket{i}|^2 + \tilde{S}_{-p(n)}[\omega_{if} + n \omega_{\rm d}] |\bra{f}\sin(\phase_j/2) \ket{i}|^2 \Big\},
\end{align}
where $\omega_{if} = (E_i - E_f)/\hbar$, and $E_i$ and $E_f$ are energies of initial and final qubit states, respectively. Parity function satisfies $p(n)= ``+"$ if $n$ is even and $``-"$ if $n$ is odd. The structure factors $\tilde{S}_\pm$ are defined in Eq.~(14) of the main text. In deriving Eq.~\eqref{eq:cp_breaking_any_states}, we approximated $Z_n(a/2) \approx 1$ which is valid to a leading order in $a$.

\bibliography{references}
\bibliographystyle{naturemag}